\documentclass[a4paper,12pt]{article}
\usepackage[utf8]{in
    putenc}
\usepackage{cancel}
\usepackage{ulem}
\usepackage{amsfonts}
\usepackage{amssymb}
\usepackage{graphicx}
\usepackage{amsmath,bm}
\usepackage{enumerate}
\usepackage{mathtools}
\usepackage{tikz} \usetikzlibrary{calc}
\usepackage[countmax]{subfloat}
\usepackage{xcolor}
\usepackage{float}
\usepackage{color}
\usepackage{here}
\usepackage{cite}
\usepackage{subcaption}
\usepackage{mwe}

\usepackage{mathrsfs}
\usepackage{float,epsfig}
\usepackage{dcolumn}
\usepackage{pgfplots}
\usepackage{graphicx}
\usepackage{bm}
\usepackage{amsmath,amssymb,amsthm}
\usepackage[colorlinks=true,linkcolor=blue,citecolor=red]{hyperref}
\definecolor{lime}{HTML}{A6CE39}
\newcommand{\orcidicon}{%
    \begin{tikzpicture}
    \draw[lime, fill=lime] (0,0)
        circle [radius=0.16]
        node[white] {{\fontfamily{qag}\selectfont \tiny ID}};
    \draw[white, fill=white] (-0.0625,0.095)
        circle [radius=0.007];
    \end{tikzpicture}   \hspace{-2mm}
}

\newcommand\orcidHasan{{\href{https://orcid.org/0000-0001-7408-0910}{\orcidicon}}}

\newcommand\orcidFaical{{\href{https://orcid.org/0000-0002-2977-0821}{\orcidicon}}}

\title{\bf On Rényi universality formula of charged flat black holes from Hawking-Page phase transition}

\author{
F. Barzi\orcidFaical\!\!$^{1,2}$\thanks{faical.barzi@edu.uiz.ac.ma},  
 H.  El Moumni\orcidHasan\!\!$^1$\thanks{h.elmoumni@uiz.ac.ma (Corresponding author)}\\
{\small $^{1}$ LPTHE, Physics Department, Faculty of Sciences, Ibn Zohr University, Agadir, Morocco. }\\
{\small $^{2}$CRMEF, Regional Center for Education and Training Professions, Marrakesh, Morocco.
}
}

\date{\today}
\begin{document} 
\maketitle
\begin{abstract}
Probing the thermodynamic properties of the Hawking-Page phase transition of asymptotically-flat black holes in Rényi statistics.
We reveal the evidence for the persisting of the dual universal formula 
associated with the Hawking-Page (HP) and minimum black hole thermodynamical transition points.  Our study unveils the universal properties of the black holes in different statistics, in particular beyond the Gibbs-Boltzmann formalism, and motivates a further promising bridge between the nonextensivity Rényi parameter $\lambda$ and the cosmological constant $\Lambda$. Besides, we discover that universal ratios related to horizon radius, Rényi temperature, and entropy can be shown to appear in presence of such a  transition. 
{\noindent}
\end{abstract}
\tableofcontents



\section{Introduction}\label{sect1}

\paragraph{}For a long time, numerous global expressions emerge in the formulation of fundamental theories. Such invariant formulas, known as universal ones, are considered independent of any coordinate system. The common type of such fundamental quantities has been of specific attention in the foundation of theories.  Their determinations are vital for the prediction of whether a theory is relevant or not, while their simple expressions as concise equations unveil profound significations and an ultimate straightforward verification.

Black holes are laboratories for testing fundamental theories that explain how the Universe works on the largest and the smallest scales (e.g., General relativity and Quantum physics) and a possible conciliation between them i.e quantum gravitational theory.
 In fact,  black hole thermodynamics plays a main role in our understanding of the mysterious nature of such compact objects \cite{Bardeen:1973gs,Hawking:1982dh}. One of the most intriguing enigmas in black hole thermodynamics is the Hawking-Page ($HP$) phase transition between the stable large black hole and the thermal gas. It was first investigated for the Schwarzschild black hole in Ref. \cite{Hawking:1982dh} and then for the charged black hole in Refs. \cite{Chamblin:1999tk,Chamblin:1999hg}, especially in the context of the AdS/CFT correspondence \cite{Witten:1998zw,Birmingham:2002ph,Herzog:2006ra,Adams:2014vza,Lala:2020lge}. After the birth of the extended phase space notion in the AdS spacetime, this area was reexamined again with prolific outcomes \cite{Spallucci:2013jja,Sahay:2017hlq,Mbarek:2018bau,Su:2019gby,Wu:2020tmz,DeBiasio:2020xkv,Li:2020khm,Belhaj:2020mdr,Yan:2021uzw,
Kubiznak:2014zwa,Kubiznak:2016qmn,Karch:2015rpa,Gregory:2019dtq,Astefanesei:2019ehu,Dehghani:2020blz,Zhao:2020nrx,Su:2021jto}. Moreover, it's largely understood that a black hole in an asymptotically flat space has negative heat capacity and is thus thermodynamically unstable. Therefore, when a black hole radiates via the Hawking mechanism \cite{Hawking:1975vcx}, it can no longer maintain thermal equilibrium with the environment and will evaporate eventually. In addition, the non-extensive nature of the Bekenstein-Hawking entropy of black holes which is proportional to the surface area of its event horizon rather than the volume, and in the strong gravitational scheme near the black hole, the condition of negligible long-range type interactions in the standard statistical descriptions breaks down, and consequently, the usual definition of mass and other extensive quantities is not possible locally pushes us to go beyond the standard Gibbs-Boltzmann ($GB$) statistical proposal.

 Alfréd Rényi in 1959 introduced a well-defined entropy function, which also obeys both the equilibrium and the zeroth law compatibility exigency of thermodynamics \cite{Renyi:1959aa}. 
\begin{equation}\label{SR}
S_R=\frac{1}{\lambda}\ln\sum_ip_i^{1-\lambda}. 
\end{equation}
In which $\lambda$ denotes a real constant parameter called the non-extensivity one.
It is obvious that in the limit $\lambda\to 0$, the standard Boltzmann-Gibbs entropy, $S_{BG}=-\sum p_i\ln p_i$ is recovered. Moreover, the Rényi entropy function  follows the non-additive composition rule 
\begin{equation}\label{tr}
S_{12}=S_1+S_2+\lambda S_1S_2, 
\end{equation}
where $S_1$, $S_2$, and $S_{12}$ are the entropies of the subsystems and the total system,
respectively. Recently,  such definition of  entropy shows exceptionally fascinating features with regards to the black holes thermodynamics and draws in a special accentuation in literature
\cite{Mejrhit:2019oyi,Tannukij:2020njz,Promsiri:2020jga,Czinner:2015eyk,Czinner:2017tjq,dilaton,Promsiri:2021hhv,Hirunsirisawat:2022ovg}. 

Through the Ref.\cite{Wei:2020kra}, in the extended phase space of the AdS spacetime, one can show that the Hawking-Page temperature and the minimum black hole one share similar dependence on the pressure, $T\sim \sqrt{P}$ with different $d$-dependent coefficients
\begin{equation}
T_{HP}=2\sqrt{\frac{d-2}{(d-1)\pi}\times P},\quad T_{min}=2\sqrt{\frac{d-3}{(d-2)\pi}\times P}.
\end{equation}
Therefrom, it's easy to conclude that $ \lim_{d\to\infty}T_{min}/T_{HP}=1$, pointed that there will be no metastable large black hole phase when $d\to\infty$. Moreover, in Ref.\cite{Wei:2020kra}, the authors discovered a fascinating  holographic-like relation between $T_{HP}(d)$ and $T_{min}(d+1)$, namely
\begin{equation}\label{mann}
T_{HP}(d)=T_{min}(d+1).
\end{equation}
Such dual relation has been extended to thermodynamics in cavity \cite{Su:2021jto}. 
%
 So far, this dual formula has not been implemented in the non-Boltzmaniann statistics. Therefore, in this paper, we will focus on the elaboration of a such holographic expression within the Rényi formalism for uncharged and charged black hole solution in flat spacetime.

\paragraph{}The outline of the paper is as follows: In Sect.\ref{sect2}, we investigate the Hawking-Page phase transition for a high-dimensional asymptotically-flat black hole in Rényi statistics and give the essential of thermodynamical quantities. In particular, we deal with the uncharged and charged solutions, and following the strategy of Ref.\cite{Wei:2020kra}, we elaborate a generalization of the dual formula associated with the anti-de-Sitter black hole in the Gibbs-Boltzmann statistics to an asymptotically-flat black hole in Rényi formalism.
Afterward, Sect.\ref{conclus} is devoted to the conclusion and open questions.
Finally, in the appendix, we investigate the thermodynamics of high-dimension charged black holes in asymptotically flat spacetime via Rényi entropy.

\section{ Dual relation associated with Hawking-Page phase portrait in Rényi formalism}\label{sect2} 
\subsection{Uncharged black hole}
\begin{figure}[H]
	\centering		
	\includegraphics[scale=.4]{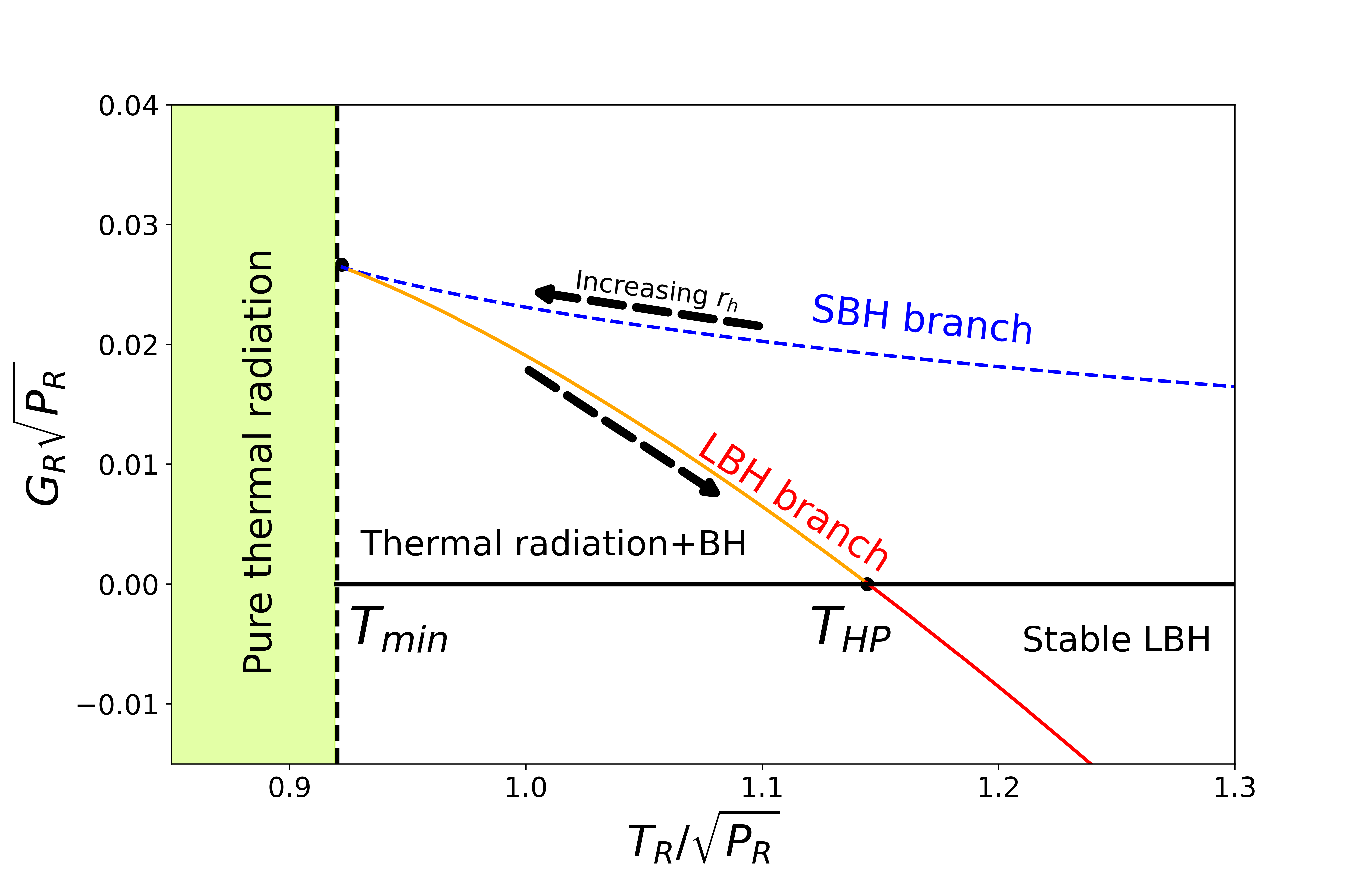}
	\vspace{-0.4cm}
	\caption{\footnotesize{\it The $G_R$-$T_R$ diagram of a four-dimensional uncharged-flat black hole in Rényi statistics. The arrows designate increasing black hole horizon radius, and $T_{HP}$ and $T_{min}$ stand for the Hawking-Page phase transition Rényi temperature and minimum Rényi temperature. The blue dashed curve is associated with the unstable small black hole branch, whereas the metastable and stable large black hole branches are depicted by the orange thin solid and red solid curves, respectively.
	}}
	\label{fig:G-T}
\end{figure}

The Hawking-Page phase transition found a deep interpretation as  confinement/deconfinement phase transition in the AdS/CFT conjecture context \cite{Witten:1998zw,Birmingham:2002ph}. While it can also be expressed as a solid/liquid phase transition within the black chemistry \cite{Kubiznak:2014zwa}, in which the cosmological constant (anti-de Sitter radius $\ell$) can be identified with the thermodynamic pressure in $d$-dimensional spacetime such as \cite{Kastor:2009wy,Belhaj:2015hha} :
\begin{equation}\label{P_ads}
P=-\frac{\Lambda}{8\pi}=\frac{(d-1)(d-2)}{16\pi \ell^2}.
\end{equation}
In the Rényi statistics, a new kind of extended phase space raises by associating now the pressure with the nonextensive parameter $\lambda$ in four dimensions as \cite{Promsiri:2020jga,dilaton} :
\begin{gather}
P_R=\frac{3\lambda}{32},
\end{gather}
and it can be put under a more general form for arbitrary spacetime dimension $d$ by following the strategy of \cite{Promsiri:2020jga}  as \footnote{The full derivation of the Rényi pressure can be found in the appendix.\ref{appendix}.}
\begin{gather}\label{P_R}
  P_R=\displaystyle \frac{  \left(d - 1\right)\left(d - 3\right) }{16 \Omega_{d} } \lambda r_{h}^{d - 4}.
\end{gather}
In order to further introduce new universal formula associated with this phase transition in the Rényi statistics approach, let us recall the behavior of the Gibbs free energy. In the extended phase space within the non-extensive parameter $\lambda$, the mass is interpreted as enthalpy, thus the Gibbs free energy can be derived as
\begin{eqnarray}\label{keyG}
G_R&=& M-T_R S_R, \\ \nonumber
&=&\frac{ r_{h}^{d - 3} (d-2)}{\Omega_{d}}-\frac{ r_{h}^{d - 5} \left(d-3\right)^2(d-1) \left(1+ \frac{64 \pi P_R r_{h}^{2}}{(d-1)(d-2)(d-3) }\right) }{64\pi \Omega_{d} P_R}\ln{\left(1+\frac{ 64 \pi P_R r_{h}^{2} }{(d-1)(d-2)(d-3)} \right)}.
\end{eqnarray}
Where, $\Omega_d$ is related to the volume of a unit $(d-2)$-sphere, $Vol(S^{d-2})$ by\cite{Zhang:2014jfa}:
\begin{equation}\label{key}
 \Omega_{d}=\frac{16\pi}{(d-2)Vol(S^{d-2})}=\frac{8 \Gamma \left(\frac{d - 1}{2}\right)}{(d - 2) \pi^{\frac{d-3}{2}}}. 
\end{equation}

and the Rényi entropy and temperature as functions of the horizon radius $r_h$ stand for
\begin{equation}\label{S_R}
S_R= 
\frac{ (d-1)(d-3) \ln{\left(1+\frac{ 64 \pi P_R r_{h}^{2} }{(d-1)(d-2)(d-3)} \right)}}{16 \Omega_{d} P_R}r_{h}^{d - 4}
\end{equation}
and
\begin{equation}\label{key}
T_R= \frac{\left(d - 3\right) \left(\Omega_{d} \left(d - 2\right) + 4 \pi \lambda r_{h}^{d - 2}\right)}{4 \pi \Omega_{d} r_{h} \left(d - 2\right)},
\end{equation}
respectively. In terms of the Rényi pressure Eq.\eqref{P_R}, the Rényi temperature becomes
\begin{equation}\label{T_R-P_R}
T_R=\frac{\left(d-3\right)}{4 \pi r_{h} }+\frac{ 16  P_{R} r_{h}}{ \left(d-2\right)\left(d-1\right)}.\end{equation}
The Gibbs free energy behavior in terms of the Rényi temperature is illustrated in the Fig.\ref{fig:G-T}
Fig.\ref{fig:G-T} clearly unveils several features: first, no black hole
solution can exist below the minimal temperature there.
\begin{equation}\label{tmin}
   T_{min} =\sqrt{\frac{16(d-3)}{\pi(d-2)(d-1)}}\sqrt{P_R},
\end{equation}
associated with the following  minimal horizon radius 
\begin{equation}
    r_{min}=\sqrt{\frac{(d-3)(d-2)(d-1)}{64\pi P_{R}}}.
\end{equation}
Second, there exist two branches above $T_{min}$, the upper one is
for small black holes which are thermodynamically unstable,
while the lower branch corresponds to a thermodynamically
stable large black holes. The positive Gibbs free energy for
$T < T_{HP}$ shows that the system is in a radiation-dominated phase.
Furthermore, The criterion of the Hawking-Page phase transition is
that the Gibbs free energy of the black hole-thermal asymptotically flat
system vanishes,
\begin{equation}\label{g=0}
 G_R=0. 
\end{equation}
We can see from Eq.\eqref{keyG} that this point
occurs at:
\begin{equation}\label{thp}
T_{HP}=\sqrt{\frac{2(d-1)}{\pi(d-2)}}\sqrt{P_R}.
\end{equation}
The radius where the Hawking-Page takes place is directly linked to the minimal radius via the following ratio,
\begin{equation}\label{gamma}
\large
\gamma_r=\frac{r_{min}}{r_{HP}}=\sqrt{\frac{d-3}{ 2}}.
\end{equation}
Such a ratio shows beautiful features, which are, that $\gamma_r$ is only dimension dependent and revealing that both Hawking-Page and minimal radii present the same behaviors in pressure $P_R$. In addition, $\gamma_r$ is an increasing function of dimension bounded from below by $\frac{\sqrt{2}}{2}$ at $d=4$.
The black hole minimum temperature Eq.\eqref{tmin} and Hawking-Page phase
transition temperature Eq.\eqref{thp} can be put under a more appropriate form as functions of the nonextensivity parameter $\lambda$
\begin{eqnarray}
T_{min}(d)&=&\displaystyle\frac{(d-2)}{4}  \left[\pi ^{\frac{d-3}{2}}\Gamma\left(\frac{d-3}{2}\right) \frac{1}{\lambda}\right]^{\frac{1}{2-d}}\label{Tmin},\\
T_{HP}(d)&=&\frac{ (d - 2)^2}{4(d - 1)}  \left[\frac{\pi ^{\frac{d-3}{2}}}{2}\Gamma\left(\frac{d-3}{2}\right) \frac{1}{\lambda}\right]^{\frac{1}{2-d}}.\label{Thp}
\end{eqnarray}
Both temperatures depend on the parameter $\lambda$ in the same manner, $T\sim\lambda^\frac{1}{d-2}$ and increase with $d$. 

\begin{figure}[!ht]
	\centering	
	\includegraphics[scale=.4]{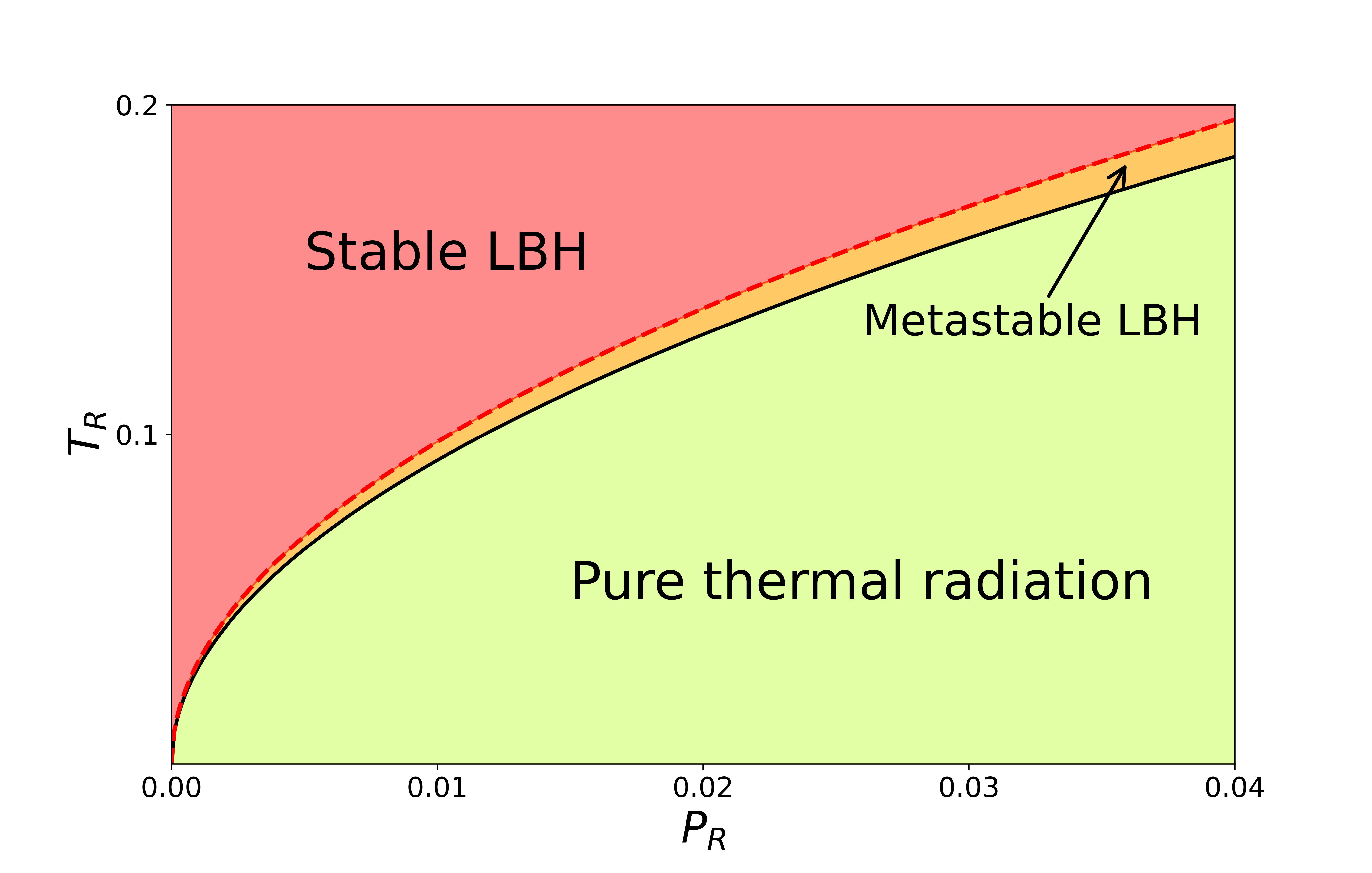}
 	\vspace{-0.5cm}
	\caption{\footnotesize{\it $T_R$-$P_R$ diagram of a four-dimensional uncharged-flat black hole in Rényi statistics. The black solid and red dashed lines extend indefinitely and represent the minimum and the Hawking-Page transition temperatures respectively at different pressures.}
	}
	\label{fig:P-T}
\end{figure}

In addition, We see from the expressions above,  \eqref{Tmin} and \eqref{Thp} that $\lambda$ fixes in $d$-dimensional spacetime a physical length scale, $l_{\lambda}=1/\lambda^\frac{1}{d-2}$ in units of Planck's length $l_p$, where the phase transition from pure thermal radiation to black-holes begins to take place. For $d=4$ the relevant length scale is $l_4=1/\sqrt{\lambda}=\sqrt{\frac{3}{32 P_R}}$ and it decreases towards a $\lambda$-independent value $l_{\lambda}=l_p$ as the dimension increases.  In addition, looking at the dependence of $T_{min}$ and $T_{HP}$ on $\lambda$ and because the non-extensive parameter is small $0<\lambda<<1$, one sees that as dimension $d$ increases it takes a smaller spacetime region filled with relatively hotter thermal radiation to form a black hole. 
Next, we depict this dependence in Fig.\ref{fig:P-T}, for $d=4$

Observing Fig.\ref{fig:P-T}, one clearly notice that three system
phases-thermal radiation, a stable large black hole, and a
metastable large black hole-are persisting. The minimum
temperature $T_{min}$ and the Hawking-Page transition temperature $T_{HP}$ are
respectively illustrated by the solid and red dashed curves. There is no terminal point in both coexistence lines, and the phase transitions can happen at all pressures, without
a critical point \cite{Kubiznak:2014zwa}. Therefore, it is more like solid/supercooled-liquid/liquid phase structure. The size of the region between the curves which represents metastable large black hole states, widens as the pressure increases. A similar behaviour in solid/liquid (thermal radiation/stable LBH) phase transitions is due to a delayed transformation from one phase to another caused by the system occupying more metastable supercooled-liquid (metastable LBH) states.

For $d=4$, we have,
\begin{eqnarray}
    T_{min}(4)&=&
    \sqrt{\frac{8}{3\pi}\times P_R},\\
    T_{HP}(4)&=&
    \sqrt{\frac{3}{\pi}\times P_R}.
\end{eqnarray}
The same behavior is observed for any higher dimension $d>4$. 
\begin{eqnarray}
T_{min}(d)&=& \sqrt{6}\sqrt{\frac{(d-3)}{(d-2)(d-1)}}\;T_{min}(4),\\
T_{HP}(d)&=&\sqrt{\frac{2}{3}}\sqrt{\frac{(d-1)}{(d-2)}}\;T_{HP}(4).
\end{eqnarray}
In addition, comparing these two quantities, we found 
\begin{equation}\label{tminthp}
\gamma_T=\frac{T_{min}(d)}{T_{HP}(d)}=\frac{4}{d-1}\sqrt{\frac{d-3}{ 2}}=\displaystyle \frac{2 \gamma_r}{\gamma_r^{2} + 1}
\end{equation}
While for the Rényi entropies we have
\begin{equation}\label{ratio_ent}
\gamma_S  = \frac{ S_{min}(d)}{ S_{HP}(d)} = \frac{ \log{\left(2 \right)}}{\log{\left(\frac{d - 1}{d - 3} \right)}}\left(\frac{d-3}{2}\right)^{\frac{d-4}{2}}=\displaystyle \frac{\gamma_r \ln{\left(2 \right)}}{2\ln(2)+\ln{\left[ \gamma_r^{2}\left( \gamma_r^{2} + 1\right) \right]}}.
\end{equation}

\begin{figure}[H]
		\centering
			\begin{tabbing}
			\centering
			\hspace{-1.4cm}
			\includegraphics[scale=.4]{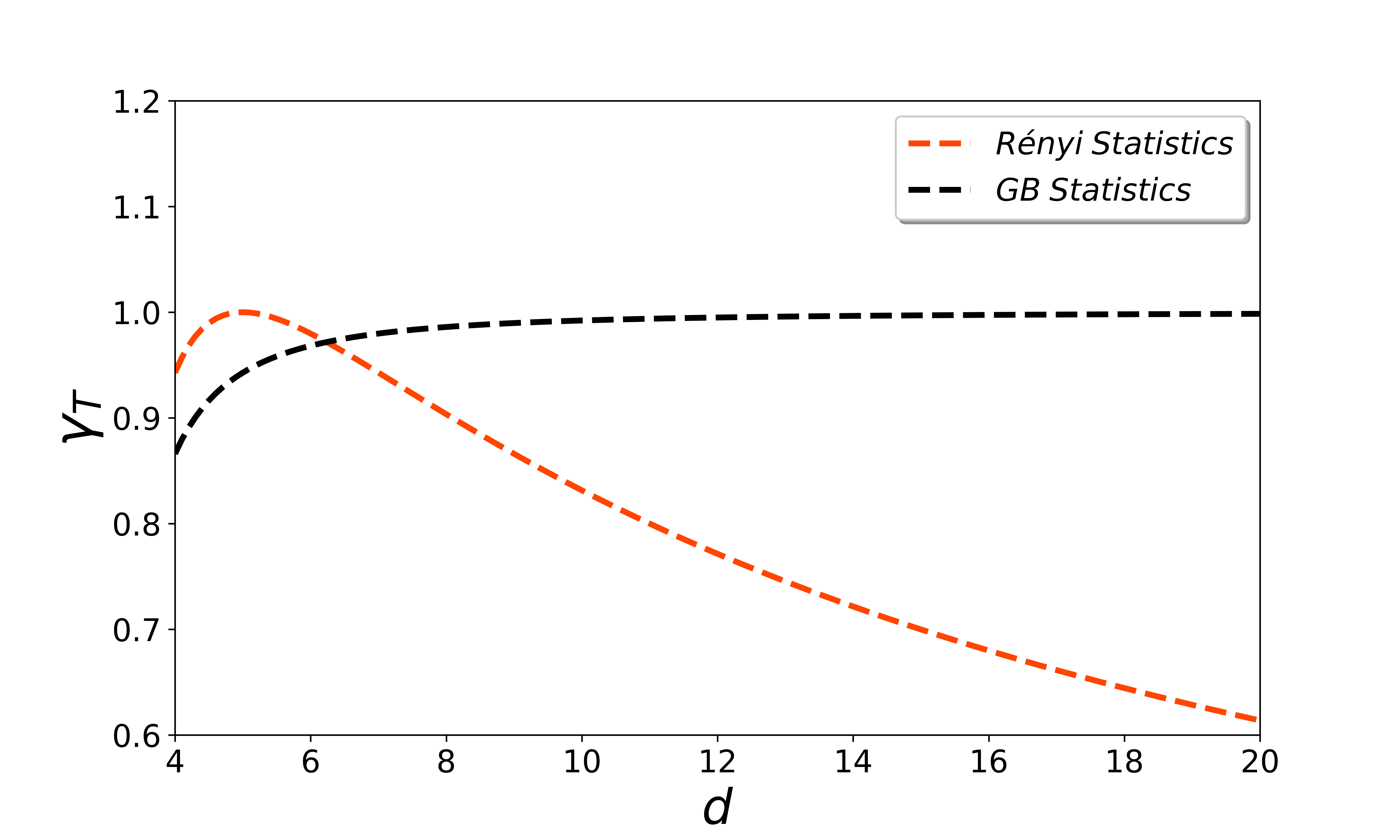} 
			\hspace{-1.6cm}
			\includegraphics[scale=.4]{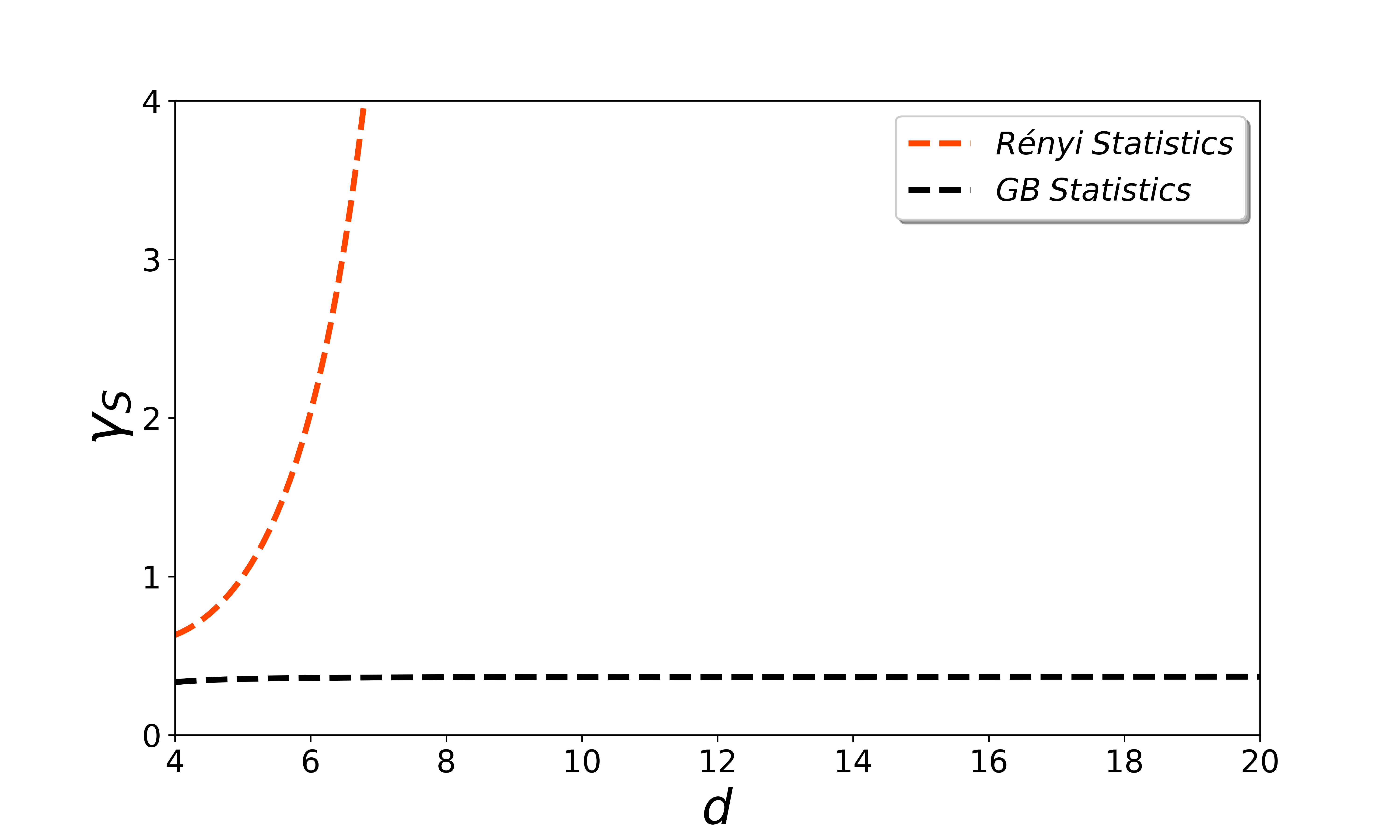} 
		   \end{tabbing}
		   \vspace{-1.cm}
 	      \caption{\footnotesize{\it Variation of temperatures and entropies ratios with spacetime dimension in Rényi and Gibbs-Boltzmann statistics (GB). In Rényi formalism, the maximum for the temperatures' ratio occurs at d=5, while for the entropies' ratio the minimum is attained at d=4. At large dimensions, $\gamma_S$ grows indefinitely but $\gamma_T$ vanishes at infinity. In Gibbs-Boltzmann formalism, however, the ratios $\gamma_T$ and $\gamma_S$ converge to $1$ and $e^{-1}$ respectively, at large dimensions. } }
  \label{fig:TminThp-d}
\end{figure}
As seen from Fig.\ref{fig:TminThp-d}, In Rényi formalism, the ratios of the two temperatures (entropies) decreases (increases) with dimension $d$, acquiring its highest (lowest) value for $d=5$ ($d=4$), as $1$ ($ \frac{\log{\left(2 \right)}}{\log{\left(3 \right)}}\approx0.6309$). This implies that in lower dimensional spacetime the collapse of thermal radiation to form a black-hole is relatively the easiest compared with higher dimensional spacetime. It is remarkable in this respect that in five dimensions once the possibility of a black hole phase exists, $T_R(5)=T_{min}(5)$, a Hawking-Page  phase transition readily happens, $T_{HP}(5)=T_{min}(5)$ and $S_{HP}(5)=S_{min}(5)$. Such a behavior is exhibited also in $GB$ formalism but only at infinite dimensions.  Furthermore, we have the limit as $ \lim_{d\to\infty}T_{min}/T_{HP}=0$  ($ \lim_{d\to\infty}S_{min}/S_{HP}=\infty $) in contrast with the anti de-Sitter space in Ref.\cite{Wei:2020kra,Su:2021jto}. This dissimilarity stems from the fact that Rényi  pressure $P_R$ depends on  the horizon radius $r_h$, Eq.\eqref{P_R}, While pressure in anti de-Sitter spacetime is independent of $r_h$, Eq.\eqref{P_ads}. 

It should be noted that in Rényi statistics, $\gamma_S$ becomes larger than unity for $d>5$, which means that the $S_{min}$ is larger than $S_{HP}$, This observation decouples the possibility of a Hawking-Page phase transition from the usual wisdom adopted within $GB$ statistics that the HP transition indicates that quantum theories of gravity ought to have a small number of states at lower energies, but a huge number of states at higher energies with a sharp transition at $T_{HP}$\cite{Belhaj:2020mdr}. Since the black hole is not an isolated system, the only criterion for such a transition is minimizing the Gibbs free energy and not maximizing the entropy. Thus it is not necessary to have $S_{min}<S_{HP}$ as predicted within $GB$ formalism.



As mentioned above, in Ref.\cite{Wei:2020kra}, the authors have established a dual relation Eq.\eqref{mann} in Gibbs-Boltzmann statistics associating the Hawking-Page temperature to minimal one of two successive dimensions. Now, we will turn our attention to checking whether it has a similar phase holographic formula in Rényi statistics formalism.

Recalling Eqs.\eqref{Tmin} and \eqref{Thp}, a straightforward calculation reveals that
\begin{equation}
 T_{min}\left(d+1,\xi_d P_R\right)= T_{hp}(d,P_R).
\end{equation}
Where we have noted $\xi_d$ as:
\begin{equation}
\xi_d=\displaystyle \frac{d}{8}\left(\frac{d - 1}{ d - 2}\right)^{2}.
\end{equation}
Therefore, we are in a position to conclude that the holographic-like relation in Eq.\eqref{mann} is a universal feature of the Hawking-Page phase transitions in different statistics and the same  interpretation remains,  like the duality between the ground and excited states of the black holes \cite{Wei:2020kra}. 

\subsection{Charged black hole}
Our holographic universal formula in Rényi statistics can be extended to the charged black hole solution.
An easy  and similar computation to the previous section reveals that the Rényi temperature calculated in Eq.\eqref{T_R-P_R} becomes in the grand canonical ensemble when the electric potential $\phi=\frac{(d-2)\Omega_d}{4(d-3)}\frac{Q}{r^{d-3}}$ remains fixed
\begin{equation}\label{TC}
T_R=\frac{(d-3)  }{4 \pi r_{h}}- \frac{ \phi^{2}(d-3)^2}{2 \pi r_{h}(d-2)}+ \frac{16 P_R r_{h}}{ (d-1)(d-2)}. 
\end{equation}
 Where the expression of the Rényi pressure is generalizes to \footnote{The full derivation of the Rényi pressure can be found in the appendix.\ref{appendix} as mentioned above.}
\begin{equation}
    P_R=\displaystyle \left[\frac{(d-1)(d-3) }{16 \Omega_{d} }-\frac{(d-1)(d-3)^2 \phi^{2}}{8 \Omega_{d} \left(d - 2\right)}\right]\lambda r_{h}^{d - 4}.
\end{equation}
While the free energy is found to be, to first order in the Rényi pressure $P_R$,  
\begin{equation}\label{GC}
G_R=\displaystyle \frac{r_{h}^{d - 3} \left[ (d-1)(d-2)-2(d-1)(d-3) \phi^{2}  - 32 \pi P_R r_{h}^{2} \right]}{\Omega_{d}(d-1)(d-2)^2}.
\end{equation}
Using the  criterion of  Eq.\eqref{g=0}, we get the expression of the Hawking-Page quantities, namely the horizon radius  
\begin{equation}\label{rhpC}
r_{HP}=\displaystyle \frac{\sqrt{2(d-1) [d-2- 2\phi^{2} \left(d-3\right)] }}{8 \sqrt{\pi} \sqrt{P_R}},
\end{equation}
and its associated temperature
\begin{equation}\label{thpC}
T_{HP}=\displaystyle \frac{\sqrt{2(d-1)}\sqrt{d-2-2\phi ^2(d-3)} }{\sqrt{ \pi }(d-2) }\sqrt{P_R}.
\end{equation}
From the above equations and like the uncharged case the dependence on $\sqrt{P_R}$ is held in the charged black hole background. Minimizing the Rényi temperature, we found the minimal horizon radius and its corresponding Rényi temperature as 
\begin{equation}\label{rminC}
r_{min}=\displaystyle \frac{\sqrt{ \left(d - 3\right) \left(d - 1\right) \left[ d - 2 - 2\phi^{2} \left(d - 3\right)\right]}}{8 \sqrt{\pi} \sqrt{P_R}},
\end{equation}
and
\begin{equation}\label{tminC}
T_{min}=\displaystyle \frac{\sqrt{16(d-3)}\sqrt{d-2-2\phi ^2(d-3)} }{\sqrt{ \pi }\sqrt{d-1}(d-2) }\sqrt{P_R}.
\end{equation}
In addition to the same dependence on the $\sqrt{P_R}$, the ratio $\gamma_r=\frac{r_{min}}{r_{HP}}$ seems to be insensible to black hole charge, for instance, it has been legitimate to consider it as a universal parameter.

In four-dimensional spacetime, the Rényi temperatures read as
\begin{equation}\label{key}
T_{min}(4)=\displaystyle \frac{2 \sqrt{6} \sqrt{P_R} \sqrt{1 - \phi^{2}}}{3 \sqrt{\pi}},
\end{equation}
and
\begin{equation}\label{key}
T_{HP}(4)=\displaystyle \frac{\sqrt{3} \sqrt{P_R} \sqrt{1 - \phi^{2}}}{\sqrt{\pi}}.
\end{equation}
By recalling, Eqs.\eqref{thpC} and  Eqs.\eqref{tminC},
the ratios of temperatures and entropies in the charged case are equal to the uncharged ones, namely,  we obtain
\begin{equation}\label{ratio_temp_RN}
 \gamma_T= \frac{T_{min}(d)}{T_{HP}(d)}=\displaystyle \frac{2 \gamma_r}{\gamma_r^{2} + 1},
\end{equation}
\begin{equation}\label{ratio_ent_RN}
 \gamma_S=  \frac{ S_{min}(d)}{ S_{HP}(d)} =\displaystyle \frac{\gamma_r \ln{\left(2 \right)}}{2\ln(2)+\ln{\left[ \gamma_r^{2}\left( \gamma_r^{2} + 1\right) \right]}}.
\end{equation}
A simple scaling arrangement  allows linking $d$-dimensional temperatures to four-dimensional ones such as,
\begin{equation}\label{key}
T_{min}(d, P_R,\phi)=T_{min}(4,\alpha_d P_R,\beta_d\phi),
\end{equation}
\begin{equation}\label{key}
T_{HP}(d, P_R,\phi)=T_{HP}(4, \Tilde{\alpha}_d P_R,\beta_d \phi).
\end{equation}
Where the scaling factors are found to be
\begin{align}
\alpha_d=\sqrt{\frac{6(d-3)}{(d-1)(d-2)}},\quad\Tilde{\alpha}_d=\sqrt{\frac{2(d-1)}{3(d-2)}}\quad \text{and}\quad \beta_d=\sqrt{\frac{2(d-3)}{d-2}}.
\end{align}
such quantities depend only on the spacetime dimension, otherwise, they
unveil a suspicion about a possible holographic-like equation due to the existence of $(d-1)$, $(d-2)$ and $(d-3)$ terms.
A close inspection unveils that the calculation is quite complicated deserving appropriate simplifications
\begin{equation}\label{holoC}
T_{min}(d+1,\xi_d P_R,\zeta_d \phi)=T_{HP}(d,P_R,\phi).
\end{equation}
Herein, we have set 
\begin{equation}\label{chiC}
\xi_d=\displaystyle \frac{d(d-1)^2}{8 (d-2)^2},\quad \text{ and }\quad \zeta_d=\displaystyle \frac{\sqrt{\left(d - 3\right) \left(d - 1\right)}}{d - 2}.
\end{equation}

At this point, it is clear that we are in presence of a dual universal relation associated with the Hawking-Page phase transition. Besides,
we remark also those other universal ratios related to horizon radius, Rényi temperature and entropy can be shown to appear in presence of such a  transition. \textcolor{magenta}{ But it is mandatory to discuss some relevant difference aspects of the dual holographic formula between the Gibbs-Boltzmann and Rényi statistics.  In particular, the holographic relation Eq.\eqref{mann} is quite simpler than the Eq.\eqref{holoC} and it is essentially due to the presence of some quantities associated with the black holes in the definition of the Rényi pressure in Eq.\eqref{P_R}, namely the event horizon radius and electric potential.  The appearance of such quantities in $P_R$ definition is because the existence of nonextensivity length $L_\lambda$, as demonstrated in appendix $B$,  and which suggests that there is a specific gravitational energy value beyond which the usual statistical mechanics is no longer valid. In other words,  the fact that when $r_h$ is greater than $L_\lambda$, the nonextensive effect appears to play a significant part in defining black hole thermodynamics. Therefore all thermodynamical quantities of the black hole become sensitive to these effects which renders the holographic relations established within non-extensive Rényi statistics relatively more complicated than those formulated in extensive conventional statistics}.

\section{Conclusion}\label{conclus}
Through this work, the dual relation between the black hole minimum temperature and Hawking-Page phase transition temperature in two successive dimensions was extended to Rényi statistics. Concretely, we reveal that both a $d$-dimensional Schwarzschild-flat and Reissner-Nordstrom-flat black holes exhibit a universal behavior at the Hawking-Page transition point. For an arbitrary dimension, the systems are characterized by two special temperatures: the Hawking-Page phase transition temperature and the black hole minimum one, which are pressure-dependent and are equal for two successive dimensions. Such equality is reminiscent of the AdS/CFT correspondence.  Rigorously speaking,  if $T_{min}$ is the temperature of a physical quantity in the bulk, thus $T_{HP}$ can be assimilated to the temperature of the dual physical quantity on the boundary. 

It is worth noting that the ratios $\gamma_r$ and $\gamma_T$ can be considered as universal quantities predicted by the charged and uncharged flat black holes. We note also that an attempt to reveal such universality in Kerr-flat black hole shows an absence of such a property, which suggests that the perfect spherical symmetry of the Schwarzschild-flat and Reissner-Nordstrom-flat black holes may be an important requirement of the observed universality. Furthermore, from the present work, this lack of universality for axisymmetric black holes is independent of the statistical formalism adopted\cite{Belhaj:2020mdr}.
\appendix
\numberwithin{equation}{section}
\section{High-dimensional flat black hole thermodynamics in Rényi formalism}\label{appendix}
In the present section, we derive the expression of the Rényi pressure $P_R$ in $d$-dimensional spacetime through the generalization of the first law of thermodynamics and the Smarr formula of asymptotically flat charged black hole in Rényi formalism. 

The starting trivial point is the line element of the $d$-dimensional Reissner-Nordstrom black-hole of mass $M$ and electric charge $Q$, in asymptotically flat spacetime is given by\cite{Zhang:2014jfa}
\begin{equation}
    ds^2=-f(r)dt^2+\frac{dr^2}{f(r)}+r^2d\omega_{d-2}.
\end{equation}
Where, the blackening function stands for 
\begin{equation}
  f(r)=1-\frac{\Omega_d M}{r^{d-3}} + \frac{\Omega_{d}^{2}  \left(d - 2\right)Q^{2} }{8 \left(d - 3\right)r^{2 d-6}},
\end{equation}
and,
\begin{equation}\label{omega_d}
 \Omega_{d}=\frac{16\pi}{(d-2)Vol(S^{d-2})}=\frac{8 \Gamma \left(\frac{d - 1}{2}\right)}{(d - 2) \pi^{\frac{d-3}{2}}}. 
\end{equation}
In the $f(r)$ function, $Vol(S^{d-2})$ and $d\omega_{d-2}$ denote the volume and the line element of the unit $(d-2)$-sphere respectively. The mass  $M$ and the electric charge $Q$, are related to the electric potential $\phi$ at the black hole outer horizon of radius $r_h$, as measured by an observer at infinity by,
\begin{align}
   M &=\displaystyle \frac{r_h^{d-3}}{ \Omega_{d}}+ \frac{2  (d - 3)\phi^{2}r_h^{d-3} }{\left(d - 2\right)\Omega_{d}}\label{BH_mass},\\
    Q &=\displaystyle \frac{4 \phi r_{h}^{d - 3} \left(d - 3\right)}{\Omega_{d} \left(d - 2\right)}\label{BH_Q}.
\end{align}
It's obvious that thermodynamical quantities at the outer horizon verify the first law of black hole thermodynamics
\begin{equation}\label{1st-law}
    dM=T_HdS_H+\phi dQ,
\end{equation}
And the Smarr formula
\begin{equation}\label{smarr}
    (d-3)M=(d-2)T_H S_H+\phi Q.
\end{equation}
Here, $T_H$ and $S_H$ are the Hawking temperature and the Hawking-Bekenstein entropy respectively, and are given by
\begin{align}
    T_H=\frac{f^{'}(r)}{4\pi},\quad S_H=\frac{r_h^{d-2}Vol(S^{d-2})}{4}.
\end{align}
In Rényi formalism, the Hawking-Bekenstein $S_H$ is considered as the Tsallis entropy and Rényi entropy $S_R$ is defined as its formal logarithm such as \cite{Promsiri:2020jga}
\begin{equation}
S_R=\frac{1}{\lambda}\ln(1+\lambda S_H),
\end{equation}
with $\lambda$ is nothing but the non-extensivity parameter. Then, Rényi temperature $T_R$ is  defined through the standard thermodynamic relation, as 
\begin{equation}
    \frac{1}{T_R}=\frac{\partial S_R}{\partial M}=\frac{1}{T_H(1+\lambda S_H)}.
\end{equation}
Now, we can readily rewrite the first law of thermodynamics Eq.\eqref{1st-law}, and the Smarr formula Eq.\eqref{smarr}, in Rényi formalism by applying the following substitutions,
\begin{equation}
    T_H=T_R\exp(-\lambda S_R),\quad S_H=\frac{\exp(\lambda S_R)-1}{\lambda}.
\end{equation}
Where $\lambda$ is taken as an infinitesimal of order one. We obtain keeping the leading order terms in $\lambda$,
\begin{align}
    dM &= T_RdS_R+VdP_R+\phi dQ,\\
    (d-3)M &= (d-2)T_R S_R-(d-2)P_RV+\phi Q.
\end{align}
In the second line, we have identified  the black hole thermodynamical volume $V$ and the Rényi pressure $P_R$ with 
\begin{align}
    V &=\displaystyle \frac{16 \pi r_{h}^{d - 1}}{\Omega_{d} \left(d - 2\right) \left(d - 1\right)},\\
    P_R &=\displaystyle \lambda r_{h}^{d - 4}\frac{(d-1)(d-3) }{16 \Omega_{d} }\left[\displaystyle1-\frac{2(d-3) \phi^{2}}{(d - 2)}\right].\label{p_lam}
\end{align}

Eq.\eqref{p_lam} permits one to express all thermodynamical quantities where $\lambda$ appears in term of the more relevant quantity, in fact, Rényi pressure $P_R$, Which leads to the following formulas used in the main text,
\begin{align}
  T_R &=\displaystyle \frac{(d-3)  }{4 \pi r_{h}}- \frac{ \phi^{2}(d-3)^2}{2 \pi r_{h}(d-2)}+ \frac{16 P_R r_{h}}{ (d-1)(d-2)}, \\
  S_R &=\displaystyle\frac{r_h^{d-4}(d-1)(d-3) \left[d-2-2 (d-3) \phi ^2\right]  \ln \left(1+\frac{64 \pi  P_R r_h^2}{(d-3) (d-1) \left(d-2-2 (d-3) \phi ^2\right)}\right)}{16 P_R(d-2) \Omega _d}.\\
  G_R &=\displaystyle \frac{r_{h}^{d - 3} \left[ (d-1)(d-2)-2(d-1)(d-3) \phi^{2}  - 32 \pi P_R r_{h}^{2} \right]}{\Omega_{d}(d-1)(d-2)^2}.
\end{align}

\numberwithin{equation}{section}
\section{Nonextensivity scale and  black hole stability requirement
}\label{appendix1}

 \textcolor{magenta}{
Over here, we investigate the effect brought up by the non-extensivity on the stability of the black hole. To this end, we consider the microcanonical ensemble of charged black holes together with
thermal radiation in asymptotically flat spacetime. The constant total energy density $E$ of the composite system is given as,
\begin{eqnarray}
E&=&E_{BH}+E_{rad}=const,\\
0&=& dE_{BH}+dE_{rad}\label{E_consrv},
\end{eqnarray}
where $E_{BH}=M-\phi Q$ and $E_{rad}=\sigma T_{rad}^4$ represent the energy density of black holes and thermal
radiation, respectively. At thermal equilibrium, the total entropy density of the system is given by, $S=S_{BH}+S_{rad}$ reaches its maximum value. Consequently, the requirements of thermal equilibrium between the black hole and its thermal environment are
\begin{equation}
\frac{\partial S}{\partial E_{BH}}=0,\quad \text{ and } \quad \frac{\partial^2 S}{\partial E_{BH}^2}<0.
\end{equation}
These two conditions serve as the stability criteria of the black hole since once the thermal equilibrium is established, the black hole can neither grow in size by absorbing energy from its surroundings nor shrink by giving energy to it. The first condition translates to
\begin{eqnarray}\nonumber
dS=0 &=& dS_{BH}+dS_{rad}\\
0 &=& \frac{dE_{BH}}{T_{BH}}+\frac{dE_{rad}}{T_{rad}}\\ \nonumber
0 &=&\frac{dE_{BH}}{T_{BH}}-\frac{dE_{BH}}{T_{rad}}
\end{eqnarray}
Where Eq.\eqref{E_consrv} is used. Thus we obtain the equality of the temperatures, $T_{BH}=T_{rad}$ at thermal equilibrium as expected. The second condition for stability gives,
\begin{eqnarray}\nonumber
\frac{\partial^2 S}{\partial E_{BH}^2}&=&\frac{\partial^2 S_{BH}}{\partial E_{BH}^2}+\frac{\partial^2 S_{rad}}{\partial E_{BH}^2}<0,\\ \nonumber
&=&\frac{\partial^2 S_{BH}}{\partial E_{BH}^2}+\frac{\partial^2 S_{rad}}{\partial E_{rad}^2}<0 \qquad \text{ since }\quad(dE_{rad}^2=dE_{BH}^2)\\
&=&-\frac{1}{T_{BH}^2}\frac{\partial^2 T_{BH}}{\partial E_{BH}^2}-\frac{1}{T_{rad}^2}\frac{\partial^2 T_{rad}}{\partial E_{rad}^2}<0,\\ \nonumber
&=&-\frac{1}{T_{BH}^2}(\frac{1}{C_{BH}}+\frac{1}{C_{rad}})<0.
\end{eqnarray}
We have therefore a condition on the heat capacities of the system
\begin{equation}\label{cond_stab2}
    \left(\frac{1}{C_{BH}}+\frac{1}{C_{rad}}\right)>0
\end{equation}
Considering the case of an infinite heat bath in thermal equilibrium with the charged black hole, the heat bath thus has an infinite number of degrees of freedom and its heat capacity becomes infinite, $C_{rad}\longrightarrow \infty$. Using Eqs.\eqref{BH_mass}, \eqref{BH_Q}, \eqref{S_R} and substituting \eqref{omega_d} and \eqref{P_R}, the inequality \eqref{cond_stab2} becomes
\begin{equation}
    \displaystyle \frac{16 \pi^{\frac{5}{2} - \frac{d}{2}} r_{h}^{8- d} \left(d - 2\right) \Gamma^{2}\left(\frac{d-3}{2} \right)\mathbf{\left[\left( d -3 \right) \pi^{\frac{d}{2} + \frac{1}{2}}\lambda r_{h}^{d-2} - 2 \pi  \Gamma\left(\frac{d-1}{2} \right)\right]} }{\left(\pi^{\frac{d}{2}} \lambda r_{h}^{d} + 2 \sqrt{\pi} r_{h}^{2} \Gamma\left(\frac{d-1}{2} \right)\right)^{2} \left(2 d \phi^{2} - d - 6 \phi^{2} + 2\right)^{2}}>0.
\end{equation}
Such constraint can be put under more elegant form as
\begin{equation}
    r_h>L_{\lambda}.
\end{equation}
Where the nonextensivity length scale $L_{\lambda}$ is defined by,
\begin{equation}
    L_{\lambda}=\pi^{\frac{1 - d}{2 \left(d - 2\right)}} \left[\frac{2}{\lambda(d-3)} \Gamma\left(\frac{d-1}{2} \right)\right]^{\frac{1}{d - 2}} 
    \end{equation}
This implies that the limit of stability of a charged black hole in asymptotically flat spacetime is fixed by the characteristic length $L_{\lambda}$ which depends only on the non-extensivity parameter $\lambda$. In energy terms, there is a particular value of gravitational energy beyond which the conventional extensive statistical mechanics is no longer valid. This is because the nonextensive effects play an essential role in describing black hole thermodynamics when $r_h$ is larger than $L_{\lambda}$.
}

\bibliographystyle{unsrt}
\bibliography{universal.bib}

\end{document}